\documentclass[11pt,twoside]{article}
\usepackage{apn3conf}

\begin{document}

\title{Lights in the Shadows, 3D-modelling knots with MOCASSIN}
\titlemark{3D-modelling knots}
\author{ Christophe Morisset}
\affil{ Instituto de Astronomia, UNAM, Mexico 
	$<$morisset@astroscu.unam.mx$>$}
\author{ Barbara Ercolano}
\affil{University College of London, UK}
\contact{Christophe Morisset}
\email{morisset@astroscu.unam.mx}
\paindex{Morisset, C.}
\aindex{Ercolano, B.}
\authormark{Morisset C. \& Ercolano B.}


\begin{abstract}
Most of the enveloppes of Planetary Nebulae (and other objects like
novae) are far from beeing homogeneous: clumps, knots and tails are
often observed. We present here the first attempt to build a
3D-photoionization model of a knot and the corresponding tail, ionized
by diffused radiation issuing from surrounding material.
\end{abstract}

\section{Introduction}

High resolution images of ionized nebulae, which are
readily obtainable with modern instruments (e.g. HST), have shown the
material in these objects to be, in the majority of cases, very
clumpy. In particular there are several small scale, high optical depth
structures associated with these nebulae; the ionization
structure in these regions appears to be very different from the 
rest of the nebular
gas. For example, images of the Helix nebula (O'Dell \& Handron, 1996;
Burkert \& O'Dell, 1998; O'Dell et al., 2003) show the presence of
numerous Knots with associated radials Tails. These knots are observed
in various PNs and may be a common situation. Even the recent images
of novae shell show that the clumpiness is more the rule than the
exeption (Bode, this conference). 
The emission of this high density knots could be partly
responsible of the t$^2$ paradigm (Peimbert, 1967). 

Van Blerkom \& Arny (1972) described theoretically the ionization of a
shadowed region illuminated by diffuse radiation coming from
surrounding ionized material. More recently, Canto et al. (1998)
presented an extension of this theoretical work and added results of
numerical gasdynamic simulations. 

We present here preliminary results of 3D photoionization models of
Helix-type knots. The main goals of such models are the understanding of the
structure and the formation of such knots and the determination of the
chemical composition of the knots and the tails. 

\section{A 3D-photoionization code is needed: MOCASSIN}

The modelling of the knots using classical 1-D photoionization codes like Cloudy 
(Ferland, 2000) is virtually impossible, since the ionization of
the tail behind the knot due to the diffusion of ionizing photons by
the surrounding material cannot be accounted for by codes based on spherical symmetry. 
The only way to model such geometry is to use a 2- or 3-D photoionization code.

The three-dimensional photoionization code, MOCASSIN
(MOnte CArlo Simulation of Ionized Nebulae, Ercolano et al., 2003 and
this conference) is used to model a cubic thick clump illuminated by
plane parallel radiation (a BlackBody at 120 kK is used for this model).

\section{Modelling knots}

\begin{figure}
\epsscale{.50}
\plotone{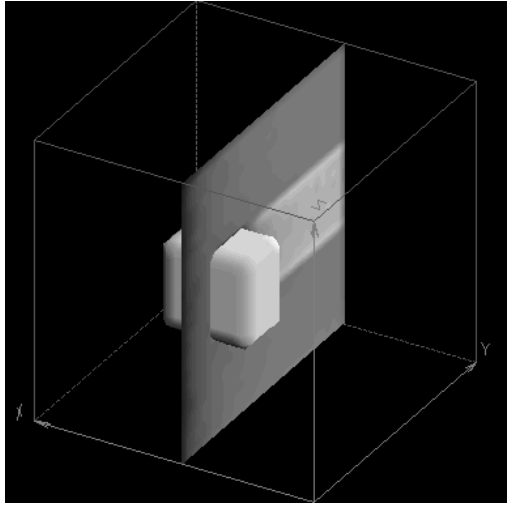}
\caption{Density distribution of the gas for the model of a Knot. The
plane paralele ionizing flux is arriving from bottom-left.} \label{fig:dens}
\end{figure}

The modeled clump has an enhanced density of 10$^5$~cm-3, while the surrounding 
gas has a density of 10$^2$~cm$^{-3}$ and
1.6~10$^2$~cm$^{-3}$ for the tail, in order to establish  pressure equilibrium with
the surrounding gas (see Fig.~\ref{fig:dens} for an illustration of
the gas distribution). 
The assumption of pressure equilibrium between the unshielded and
shadowed regions leads to an increase of the density in the shadowed
region (the electron temperature being lower). The emissivity of
recombination lines will then be higher in the shadow. 
The ionization state of the shadowed regions will be lower than in the
surrounding material, which is directly illuminated by the primary radiation field, 
resulting in the enhanced emission from low charged ions from these regions. 

A grid of 30$^3$ cells is used. The results of the model
presented in the next section are preliminary and should only be
considered as a test case with the scope to demonstrate the
suitability of the MOCASSIN code to this kind of environment, rather
than an attempt to reproduce all the observed characteristics of the
Helix Knots.  

\section{Prelimininary results}

\begin{figure}
\epsscale{0.80}
\plotone{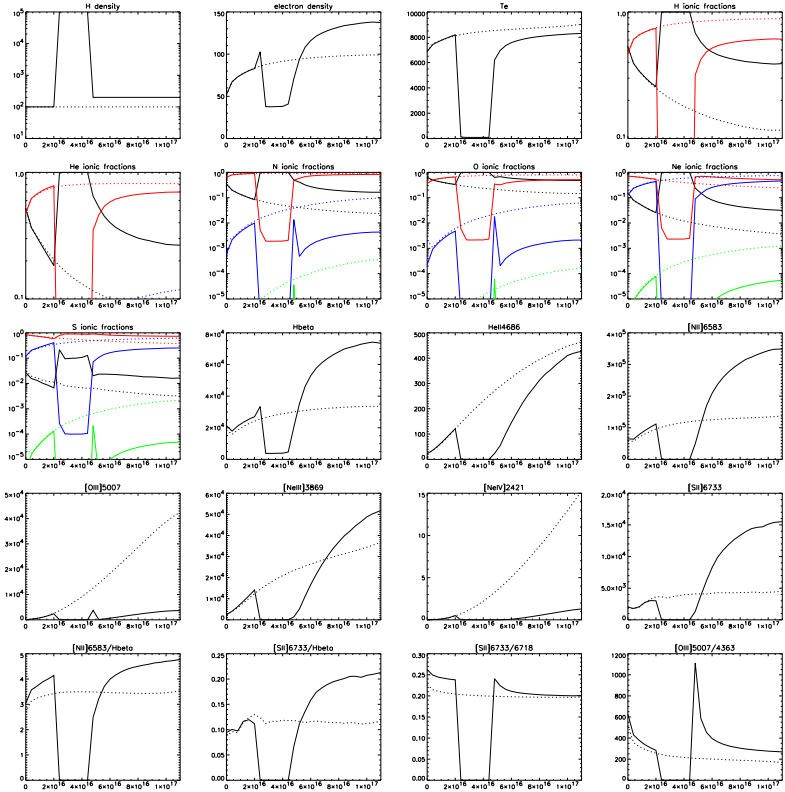}
\caption{Results of MOCASSIN model, in the Y-axis direction, along a line
crossing the dense clump (solid lines)  and following a line through
the surrounding directly photoionized region (dashed lines). Electron
temperature and pressure, ionic fractions (black, red, blue and green
for 0, 1, 2, 3 times ionized resp., color version available on
astro-ph/0310***) and line emissions are presented. 
} \label{fig:ionifrac}
\end{figure}

\begin{figure}
\epsscale{.50}
\plotone{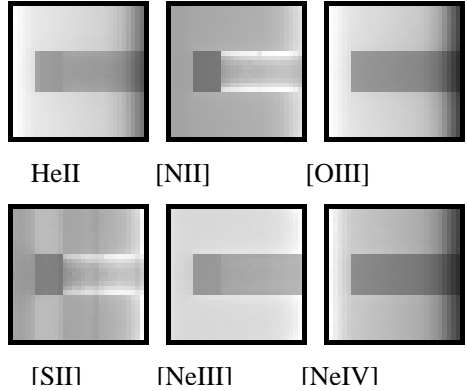}
\caption{Synthetic images in the sky plane, for some emission lines. The main
trends observed in the Helix Knots are reproduced : recombination and
forbidden lines from low charged ions are emitted preferentially by
the tail.  
} \label{fig:images}
\end{figure}

Fig.~\ref{fig:ionifrac} shows the results of the ionic fraction
distribution and proves that the tail of the knot is effectively
ionized by the diffuse radiation. Fig.~\ref{fig:images} shows the
image obtained for various emission lines, the results agree quite
well with the Helix observations. Nevertheless, the contrast of
emission between the tail and the surrounding material we obtain here
only reproduce the observations because the geometrical depth of the
surrounding material is small. If a more realistic geometry were to be used
(considering the geometrical size of the tail compared to the
surrounding gas), the enhanced emission of the tail would vanish; the
hypothesis of pressure equilibrium seems then not to be valid.

\section{Conclusion and future works}

MOCASSIN is able to deal with shadows ionized by diffused
radiation. The first results presented here are promising, and more 
realistic models will be performed to reproduce all the observational
constraints obtained for the Helix knots. In particular, the
hypothesis of pressure equilibrium between the tail and the
surrounding gas has to be carefully investigated.

\end{document}